\documentclass[preprint2]{aastex}
\usepackage{natbib,longtable,hyperref}

\newcommand{\beq}{\begin{equation}}
\newcommand{\eeq}{\end{equation}}

\slugcomment{Accepted version 1.1}

\begin{document}
 \DeclareGraphicsExtensions{.pdf,.gif,.jpg}

\title{NEOWISE: Observations of the Irregular Satellites of Jupiter and Saturn}

\shorttitle{NEOWISE: Observations of Irregular Satellites}
\shortauthors{Grav {\it et al.}}
\medskip

\author{T.~Grav} 
\affil{Planetary Science Institute, Tucson, AZ 85719, USA; tgrav@psi.edu}
\author{J.~M.~Bauer\altaffilmark{1}, A.~K.~Mainzer, J.~R.~Masiero} 
\affil{Jet Propulsion Laboratory, California Institute of Technology, Pasadena, CA 91109, USA}
\altaffiltext{1}{Infrared Processing and Analysis Center, California Institute of Technology, Pasadena, CA 91125, USA}
\author{ C.~R.~Nugent, R.~M.~Cutri} 
\affil{Infrared Processing and Analysis Center, California Institute of Technology, Pasadena, CA 91125, USA}
\author{S.~Sonnett, E.~Kramer} 
\affil{Jet Propulsion Laboratory, California Institute of Technology, Pasadena, CA 91109, USA}

\date{\rule{0mm}{0mm}}
\begin{abstract}
We present thermal model fits for 11 Jovian and 3 Saturnian irregular satellites based on measurements from the WISE/NEOWISE dataset. Our fits confirm spacecraft-measured diameters for the objects with {\it in situ} observations (Himalia and Phoebe) and provide diameters and albedo for 12 previously unmeasured objects, 10 Jovian and 2 Saturnian irregular satellites. The best-fit thermal model beaming parameters are comparable to what is observed for other small bodies in the outer Solar System, while the visible, W1, and W2 albedos trace the taxonomic classifications previously established in the literature. Reflectance properties for the irregular satellites measured are similar to the Jovian Trojan and Hilda Populations, implying common origins. 
\end{abstract}
\keywords{planet and satellites: detection - planet and satellites: fundamental parameters- infrared: planetary systems - surveys}
\section{Introduction}

The irregular satellites are natural moons of the giant planets in our solar system. While the {\it regular} satellites, like the Galilean are on prograde, near-circular orbits close to the equatorial plane of their host planet, the {\it irregular} satellites are on more distant, highly eccentric and highly inclined orbits. A major fraction of the irregular satellites are retrograde with inclinations greater than $90^\circ$. Currently there are 107 known irregular satellites (54 orbit Jupiter; 38 orbit Saturn; 9 orbit Uranus; and 6 orbit Neptune). The difference in the known numbers is generally attributed to the increasing difficulty experienced in observing these objects as the distances to the planets increase. 

One major feature of the irregular satellites is that their orbits are dynamically clustered into families \citep{Gladman.2001a, Sheppard.2003a, Nesvorny.2003a}. The families orbiting Jupiter have been named after their largest known member. Jupiter has at least five known groups, two of which are prograde (with the biggest members being Himalia and Themisto) and three that are retrograde (Ananke, Carme and Pasiphae). The families of irregular satellites around Saturn have been named for the mythos that inspired the names of their members. There are two known prograde groups, the Inuit and the Gallic. The retrograde satellites, except for Phoebe, are named after nordic ice giants, and no clear family structure is apparent, although more than three different groupings may exist. While the processes that created some of the families (like the Carme and Ananke familes at Jupiter) are known to be collisional in origin, others (like the inclination groups at Saturn) remain unknown. 

The irregular satellites are believed to have been heliocentric asteroids that were captured during the early stages of solar system formation \citep{Colombo.1971a,Heppenheimer.1977a,Pollack.1979a,Agnor.2006a, Nesvorny.2007a,Nesvorny.2014a}. Understanding when and how these bodies were captured can provide constraints on and understanding of how the giant planets, and thus the entire solar system, formed. Determining the source regions of the captured asteroids also provides an important window into the physical and dynamical evolution of these populations. It was recognized early that the current population of irregular satellites might have resulted from of a small number of captures with subsequent fracturing events \citep{Pollack.1979a}, rather than a larger number of individual captures \citep{Bailey.1971a,Bailey.1971b}. 

The optical and near-infrared color distribution of the irregular satellites show a range from neutral/grey (Sun-colored) to moderately red colors \citep{Smith.1981a,Tholen.1984a,Luu.1991a,Rettig.2001a,Grav.2003b, Grav.2007a}, consistent with C-, P- and D-type asteroids. \citet{Grav.2003b} found that the colors within a dynamically defined family are more similar to each other than members of other families. This indicates that the irregular satellites from a dynamical family are indeed fragments from a single, homogeneous and larger progenitor. It is, however, possible that surface processes, like space weathering, may lead to the observed uniformity \citep{Jewitt.2007a}. 

The origins of the irregular satellites remain unknown. \citet{Hartmann.1987a} suggested that since most of the large irregular satellites have C-type spectra they were objects in the outer main asteroid belt ejected due to Jupiter resonances, such as the Kirkwood gaps near the 5:2 and 7:3 resonances. \citet{Johnson.2005a} used the density derived from the close fly-by of S9 Phoebe by the Cassini-Huygens spacecraft on 11 June 2004 to suggest that this irregular satellite originated in the outer solar system, rather than {\it in situ} in the Saturnian system.  The colors of the irregular satellites of Jupiter and Saturn are systematically bluer than the colors of the Centaur and Trans-Neptunian populations, and they lack the {\it ultrared} surfaces found in these two populations \citep{Jewitt.2002a,Jewitt.2007a}. This difference in color distribution could be due to different dynamical, physical or collisional histories, but could also signify that the Centaur population (and thus the trans-Neptunian population) may not be the source region for the irregular satellites. 

If this color difference remains as more optical and near-infrared photometry of the irregular satellites are collected, it sets an important constraint on the source regions from which the irregular satellites may possibly have been captured. It is noted that this difference in colors between the irregular satellites and the Centaur population are due to different dynamical histories or different collisional histories 

Some caution should be taken here with regards to the use of C-, P- or D-type classification for the irregular satellites. These spectral classifications are routinely used to describe the surfaces of the outer main belt asteroids and cometary objects in the outer disk. The terminology more commonly used for the more distant Centaur and trans-Neptunian populations are neutral/grey, red and ultra-red to describe their spectral slopes. Since the origins of the irregular satellites are unknown, but could be either from the outer main belt, the trans-Neptunian population (through the Centaur population), or a combination of the two it can be hard to choose what nomenclature to use in describing their observed or derived spectral slopes. In this paper we use the C-, P- or D- terminology, in order to be consistent with most of the historical papers and our own series of papers based on the WISE data. It should be cautioned that this use does not imply any additional knowledge, beyond the fact that the spectral slope of the observed irregular satellites have similar spectral slopes to that of the C-, P- and D-type asteroids.

A more complete review of the current understanding of the irregular satellites' physical and dynamical properties, origins and evolution can be found in \citet{Jewitt.2007a} and \citet{Nicholson.2008a}. In this paper we present the observation of three Saturnian and eleven Jovian irregular satellites with WISE as well as the thermal modeling performed to derive their effective diameters and albedos. 

\section{Observations} 

WISE is a NASA Medium-call Explorer mission which during 2010 surveyed the entire sky in four infrared wavelengths, $3.4$, $4.6$, $12$ and $22\mu$m \citep[denoted W1, W2, W3, and W4, respectively;][]{Wright.2010a,Mainzer.2005a}. The solar system-specific portion of the WISE project, known as NEOWISE, collected observations of more than 158,000 asteroids, including near-Earth objects (NEOs), main belt asteroids (MBAs), comets, Hildas, Jovian Trojans, Centaurs, and scattered disk objects \citep{Mainzer.2011a}. Both the WISE and NEOWISE portions of the survey and instructions on retrieval of data from the WISE databases are described in complete detail in \citet{Mainzer.2011a} and \citet{Grav.2011b}. 

We found observations of nine irregular satellites of Jupiter in the WISE single-exposure image and extract source data. Five of the objects (J6 Himalia, J7 Elara, J8 Pasiphae, J9 Sinope, and J11 Carme, ) were detected in all four bands. One object (J10 Lysithea) was detected in W1, W3 and W4. The remaining three objects (J12 Ananke, J13 Leda, and J17 Callirrhoe) were detected in the two longest bands, W3 and W4. While J18 Themisto should be bright enough to be seen by WISE, it crossed the spacecraft field-of-view while it was close to Jupiter and its flux was completely drowned out by the scattered light from the planet. Observations of two additional Jovian irregular satellites, J23 Kalyke and J27 Praxidike, were extracted in W3 by stacking all available observations of the objects (see Section \ref{sec:stack}). Stacks were also created using the ephemerides of J19 Megaclite, J20 Taygete and J24 Iocaste, but no signal was detected for these objects. 

Only S9 Phoebe was bright enough among the Saturnian irregular satellites to be detected by WISE in single observations, with strong signal available in the W1, W3 and W4 bands.  We were able to use stacking (see Section \ref{sec:stack}) to extract W4 observation of S26 Albiorix and S29 Siarnaq. Stacks were attempted for S20 Palliaq and S21 Tarvos, but no signal was detected in either band for these two objects.  

\begin{deluxetable}{lrrcccccccc}
\tablecaption{Observations: The observational circumstances, number of observations and published optical magnitudes are given. Note that for all the Jovian irregular satellites observed here, the absolute magnitude is found using an assumed value of IAU phase slope parameter $G =0.10\pm0.15$ (see section \ref{sec:thermalmodel} for more detail). The note $^*$ indicates that the observations were stacked.}
\tablehead{
\colhead{Name} & \colhead{H}  & \colhead{G} & \colhead{MJD} & \colhead{r} &
\colhead{$\Delta$} & \colhead{$\alpha$} & \multicolumn{4}{c}{Number of Observations} \\
 & & & & & & & \colhead{W1} & \colhead{W2} & \colhead{W3} & \colhead{W4} \\ 
 & \colhead{[mag]} && \colhead{[UTC]} & \colhead{[AU]} & \colhead{[AU]} & \colhead{[deg]}
}
\startdata
J6 Himalia & $8.00\pm0.01$\tablenotemark{1} & $0.10\pm0.15$ & 55369 & 4.903 & 4.812 &  11.96 & 15 & 15 & 14 & 15 \\
 	& && 55540 & 4.953 & 4.746 & 11.40 & 12 & 12 & &  \\
J7 Elara & $9.45\pm0.01$\tablenotemark{1} & $0.10\pm0.15$ & 55370 & 4.912 & 4.823 & 11.94 & 13 & 11 & 13 & 13 \\
	& && 55540 & 5.019 & 4.810 & 11.24 & 13 & 13 & & \\
J8 Pasiphae & $10.21\pm0.01$\tablenotemark{1} & $0.10\pm0.15$ & 55368 & 4.877 & 4.784 & 12.03 & 14 &  & 13 & 14\\
	& && 55540 & 5.023 & 4.814 & 11.23 & 11 & \\
J9 Sinope & $11.29\pm0.01$\tablenotemark{1} &$0.10\pm0.15$ & 55371 & 4.819 &  4.728 & 12.17 & 13 & & 14 & 14 \\
J10 Lysithea & $11.09\pm0.02$\tablenotemark{1} & $0.10\pm0.15$ & 55368 & 5.011 & 4.920 & 11.70 & 8 & & 15 & 15\\
J11 Carme & $10.91\pm0.01$\tablenotemark{1} &$0.10\pm0.15$ & 55369 & 4.831 & 4.738 & 12.14 & 16 & 9 & 16 & 16 \\
	& & & 55540 & 5.062 & 4.858 & 11.15 & 8 & 6 & & \\
J12 Ananke & $11.84\pm0.03$\tablenotemark{1} &$0.10\pm0.15$ & 55370 & 5.058 & 4.971 & 11.59 & & & 15 & 15 \\
J13 Leda & $12.63\pm0.03$\tablenotemark{1} & $0.10\pm0.15$ & 55368 & 5.038 & 4.948 & 11.64 & & & 12 & 12\\
J17 Callirrhoe & $13.92\pm0.02$\tablenotemark{2}  & $0.10\pm0.15$ & 55371 & 4.877 & 4.788 & 12.03 & & & 10 & 8 \\
J23 Kalyke & $15.28\pm0.04$\tablenotemark{2}  & $0.10\pm0.15$ & 55367 & 4.965 & 4.872 & 11.81 & & & $14^*$\\
J27 Praxidike & $15.24\pm0.03$\tablenotemark{2}  & $0.10\pm0.15$ & 55367 & 5.038 & 4.946 & 11.63 & & & $13^*$\\
S9 Phoebe & $6.59\pm0.02$\tablenotemark{3} & $0.02\pm0.03$\tablenotemark{3} & 55360 & 9.581 & 9.419 & 6.05 & \\
S26 Albiorix & $11.35\pm0.05$\tablenotemark{3} & $0.39\pm0.06$\tablenotemark{3} & 55359 & 9.500 & 9.338 & 6.10 &&&& $11^*$\\
S29 Siarnaq & $10.90\pm0.05$\tablenotemark{3} & $0.45\pm0.17$\tablenotemark{3} & 55359 & 9.410 & 9.247 & 6.16 &&&& $13^*$\\
\enddata
\tablerefs{1) \citet{Rettig.2001a}; 2) \citet{Grav.2003b}; 3) \citet{Bauer.2006a}}
\end{deluxetable}

\subsection{Image Stacking}
\label{sec:stack}

For two of the Jovian and two of the Saturnian irregular satellites it was possible to perform a shift and co-added image stack procedure to increase detection sensitivity, that allowed for identification of thermal detections that were not found in the individual images. This method of shift and stack has been done with great success on WISE detections of comets, Centaurs and scattered disc objects \citep{Bauer.2011a,Bauer.2012a,Bauer.2012b,Bauer.2013a}. Images containing the predicted positions of the objects were identified using the IRSA/WISE image server solar system object search tool as described by \citet{Cutri.2012a}. The images were co-added using "A WISE Astronomical Image Co-adder" \citep[AWAIC;][]{Masci.2009a}. This stacking resulted in $\sim3-4\sigma$ detections of J23 Kalyke and J27 Praxidike in the W3 band, and $\sim 3-6\sigma$ detections of S26 Albiorix and S29 Siarnaq. The extracted magnitudes are tabulated in Table \ref{tab:stack}. 

\begin{table}[htdp]
\caption{Photometry from Co-added Images}
\begin{center}
\begin{tabular}{lrrr}
Name & MJD & W3 & W4  \\
 & [UTC] & [mag] & [mag] \\
\hline
J23 Kalyke      & 55367 & $11.32\pm0.24$ & \\
J27 Praxidike & 55367 & $11.43\pm0.28$ & \\
S26 Albiorix & 55359 & & $7.95\pm0.29$ \\
S29 Siarnaq & 55359 & &  $7.22\pm0.16$ \\
\hline
\end{tabular}
\label{tab:stack}
\end{center}
\end{table}%

\section{Thermal Modeling}
\label{sec:thermalmodel}

In this paper, we use the thermal modeling methods described in our previous papers on WISE detections of the Hilda and Jovian Trojan populations \citep{Grav.2011b,Grav.2012a,Grav.2012b}. Absolute magnitudes were collected from \citet{Luu.1991a}, \citet{Rettig.2001a}, \citet{Grav.2003b,Grav.2007a} and \citet{Bauer.2006a}, thus allowing for the derivation of visible albedos. The errors of these absolute magnitudes are based on the photometric accuracy, but generally uses an assumed phase coefficient of $G = 0.15$, which is commonly used for small bodies for which the phase function behavior has not been derived. \citet{Bauer.2006a} did derive G values in the R and B band for 6 Saturnian irregular satellites, including the three detected by WISE. The values ranged from $-0.11\pm0.17$ (for S22 Ijiraq) to $+0.45\pm0.17$ (for S29 Siarnaq). The inverse variance weighted mean G value for the Saturnian irregular satellites is $0.10\pm0.15$, which we adopt as the value for objects where the phase function is not available; none of the Jovian irregular satellites have measured G values. The phase functions can now be used to estimate the error in H due to assuming a standard value of $G=0.15$.  Most of the photometric measurements of the irregular satellites are done at 2 to 12 degrees phase angle, which means that a difference in G of $\pm 0.15$ would yield errors in H of $0.04$ and $0.15$ magnitudes for the two bounding phase angles. We thus conservatively set the error on H at $0.15$ magnitudes for all objects for which a phase function has not been derived.

\section{Results}

The detected Jovian irregular satellites range in sizes from $139.6\pm1.7$km in diameter for J6 Himalia to as small as $6.9\pm1.3$km for J23 Kalyke and $7.0\pm0.7$km for J27 Praxidike. J17 Callirrhoe represents the smallest Jovian irregular satellite seen in individual images with a diameter of $9.6\pm1.3$km. For the Saturnian irregular satellites the results are consistent with the size derived from the flyby of S9 Phoebe \citep{Porco.2005a} at the $2\sigma$ level. The diameters of S26 Albiorix and S29 Siarnaq were found to be $28.6\pm6.3$km and $39.3\pm5.9$km, respectively. 

\begin{figure}[h]
\begin{center}
\includegraphics[width=8cm]{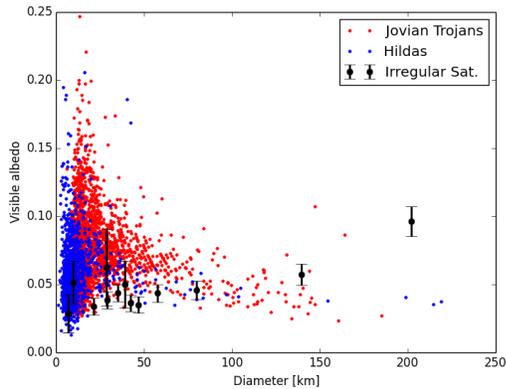}
\caption{The visible albedo distribution of the irregular satellites detected with WISE as a function of diameter. The observed population of Jovian irregular satellites is darker than both the Jovian Trojan and Hilda populations \citep{Grav.2011b,Grav.2012a}.}
\label{fig:DvspV}
\end{center}
\end{figure}

All the objects detected by WISE are very dark, with visible geometric albedos ranging from $2.6$ to $9.6\%$, consistent with those of the C-, P- and D-type asteroids that dominate in the region between the main asteroid belt and the giant planets. Figure \ref{fig:DvspV} shows the derived diameter and visible albedo of the irregular satellites from the thermal modeling. The results are compared to that of the Hilda population \citep[blue points;][]{Grav.2012a} and the Jovian Trojan population \citep[red points;][]{Grav.2011b,Grav.2012b}. The largest Saturnian irregular satellite, S9 Phoebe, is by far the brightest of the irregular satellites observed, with a visible albedo of $9.6\pm1.7\%$. 

The mean weighted albedo of the 14 irregular satellites observed in this paper is $4.3\pm1.4$\%, which makes the irregular satellites one of the darkest known populations in the solar system. The inverse variance weighted mean albedo of the Jovian sample is slightly darker at $4.0\pm0.8$\%. The irregular satellites are as dark as both the Hilda population \citep[which has a mean weighted visible albedo of $5.5\pm1.8\%$;][]{Grav.2012a} and the Jovian Trojan population \citep[with a mean weighted visible albedo of $7\pm3$][]{Grav.2011b}. It is also as dark as the dark sub-population of the Main Belt Asteroids, which were found to have a mean albedo of $6\pm3\%$ \citep{Masiero.2011a}. The blue component of the Centaur population has a mean weighted albedo of $6\pm2$ \citep{Bauer.2013a} and would provide the connection to the suggested outer solar system as a possible source region. However, the connection to the Centaur population raises a problem of explaining why only members of the blue component of Centaurs were captured, since the irregular satellites lack any members with sufficiently red colors or high enough albedo to be consistent with captures from the red component of the Centaur population. Alternatively all captured Centaurs would have to have undergone some dynamical or collisional process (for example heating due to low perihelia passages, space weathering, or exposing of new ice) before they were captured as irregular satellites. The irregular satellites rival the comet population as the on of the darkest populations in our solar system \citep{Lamy.2004a,AHearn.2005a,AHearn.2011a,Capaccioni.2015a}. 

\begin{figure}[h]
\begin{center}
\includegraphics[width=8cm]{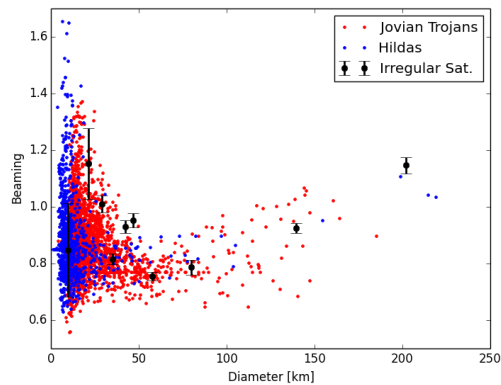}
\caption{The beaming parameter distribution of the irregular satellites is shown as a function of diameter and compared to the beaming parameter distribution for the Hilda and Jovian Trojan populations.}
\label{fig:Dvseta}
\end{center}
\end{figure}

The beaming values are more scattered, ranging from $0.76\pm0.02$ for J8 Pasiphae to $1.15\pm0.10$ for J13 Leda. Figure \ref{fig:Dvseta} shows the derived diameters and beaming values for the 10 irregular satellites for which the beaming parameter could be derived. The results are again compared to the Hilda population \citep[blue points;][]{Grav.2012a} and the Jovian Trojans \citep[red points;][]{Grav.2011b,Grav.2012b}; most of the irregular satellite have beaming values comparable to these two populations. For the four stacked objects with detections in a single band we adopt a default beaming value of $0.9\pm0.2$. This is consistent with the beaming values found among the Hilda population, Jovian Trojan population, and blue component of the Centaur population. A smooth, spherical asteroid with no thermal intertia has a beaming value of $1$ \citep{Harris.2002a}. A lower value than $1$ is generally thought to be caused by surface roughness, while high thermal inertia and rotation cause an increase. All of the irregular satellites detected by WISE are consistent with low thermal inertia. Most of the large Jovian irregular satellites have beaming values that are consistent with some surface roughness. 
 
\begin{deluxetable}{lrrrrrcccc}
\tablecaption{Thermal Fit Results: The diameter, visible and infrared albedo, and beaming parameter derived from the thermal modeling, together with the number of observations used in each band are given. The note $^*$ indicates that the observations were stacked.}
\tablehead{
\colhead{Name} & \colhead{Diameter} &
\colhead{Beaming} & \multicolumn{3}{c}{Albedo}  & \multicolumn{4}{c}{Number of Observations} \\
 & && \colhead{Visible} & \colhead{W1} & \colhead{W2} & \colhead{W1} & \colhead{W2} & \colhead{W3} & \colhead{W4}  \\
 &\colhead{[km]}  & & \colhead{\%} &  \colhead{\%} & \colhead{\%} \\
}
\startdata
J6 Himalia  & $139.6\pm1.7$ & $0.93\pm0.02$ & $5.7\pm0.8$ & $7.0\pm0.7$ & $6.4\pm0.6$ & 26 & 26 & 14 & 15 \\ 
J7 Elara &         $79.9\pm1.7$ & $0.79\pm0.03$ & $4.6\pm0.7$ & $6.3\pm0.5$ & $5.2\pm1.1$ & 26 & 24 & 13 & 13 \\
J8 Pasiphae & $57.8\pm0.8$ & $0.76\pm0.02$ & $4.4\pm0.6$ & $6.7\pm0.7$ & & 25 & & 13 & 14 \\
J9 Sinope & $35.0\pm0.6$ & $0.82\pm0.02$ & $4.2\pm0.6$ & $10.8\pm1.2$ & & 13 &  & 14 & 13 \\
J10 Lysithea & $42.2\pm0.7$ & $0.93\pm0.02$ & $3.6\pm0.6$ & $6.9\pm1.1$ & & 8 & & 15 & 15 \\
J11 Carme & $46.7\pm0.9$ & $0.95\pm0.03$ & $3.5\pm0.6$ & $9.7\pm1.0$ & $18.5\pm2.2$ & 24 & 15 & 16 & 16 \\
J12 Ananke & $29.1\pm0.6$ & $1.01\pm0.03$ & $3.8\pm0.6$ & & & & & 15 & 15 \\
J13 Leda & $21.5\pm1.7$ & $1.15\pm0.13$ & $3.4\pm0.6$ & & & & & 12 & 12\\
J17 Callirrhoe & $9.6\pm1.3$ & $0.85\pm0.17$ & $5.2\pm1.6$ & & & & & 10 & 8 \\
J23 Kalyke & $6.9\pm1.3$ & & $2.9\pm1.4$ & & & & & $14^*$ & \\
J27 Praxidike & $7.0\pm0.7$ &  & $2.9\pm0.6$ & & & & & $13^*$ &\\
S9 Phoebe & $202.2\pm4.5$ & $1.15\pm0.03$ & $10.0\pm0.5$ & $5.5\pm0.5$ & $9.8\pm1.6$ & 10 & 7 & 11 & 11 \\
		& $213.2\pm2.0$ & $1.22\pm0.02$ & $9.0\pm0.2$ & $5.0\pm0.4$ & $8.8\pm1.3$ & 10 & 7 & 11 & 11 \\
S26 Albiorix & $28.6\pm5.4$ & & $6.2\pm2.8$ & & & & & & $11^*$ \\
S29 Siarnaq & $39.3\pm5.9$ && $5.0\pm1.7$ & & & & & & $13^*$ \\
\enddata
\end{deluxetable}


\subsection{J6 Himalia} 
J6 Himalia is the largest of the Jovian irregular satellites, and it is, except for Phoebe, the most extensively observed of all the irregular satellites. Himalia was discovered in 1904 by Charles~D.~Perrine using the Lick observatory. It orbits Jupiter in a prograde orbit with mean inclination of $\sim28^\circ$ and a semi-major axis of $~11.5$ million km (or $\sim165$ Jupiter radii). \citet{Cruikshank.1977a} was the first to derive a diameter, $170\pm20$km, and albedo, $3\%$, based on radiometric measurements in the $20\mu$m band using the University of Hawaii 2.2m telescope on Mauna Kea, Hawaii. In late 2000, Himalia was observed by the \emph{Cassini} spacecraft resulting in disk-resolved images \citep{Porco.2003a}. With the spacecraft 4.4 million kilometers away from the irregular satellite, the image scale was $27$km/pixel. Himalia subtended 4 to 6 pixels, indicating an ellipsoid shape corresponding to a size of $150\pm20$km by $120\pm20$km. \citep{Porco.2003a} combined this new size and ground based photometry to determine an updated albedo value of $5\pm1\%$.

There were 15 detections of Himalia in the WISE dataset, with flux measurements in all 4 bands. Thermal modeling of Himalia yields an effective diameter of $139.7\pm1.8$km and a visible albedo of $5.7\pm1.2$\%, which are consistent with the earlier measurements. With a W1 band albedo of $7.0\pm0.7\%$ \citep[cf][]{Grav.2012b}, Himalia is consistent with a C-type taxonomy as reported in \citet{Tholen.1984a,Rettig.2001a}. A beaming value of $0.93\pm0.02$ indicates that Himalia has low thermal inertia, with some surface roughness. 

\begin{figure}[ht]
\begin{center}
\includegraphics[width=8cm]{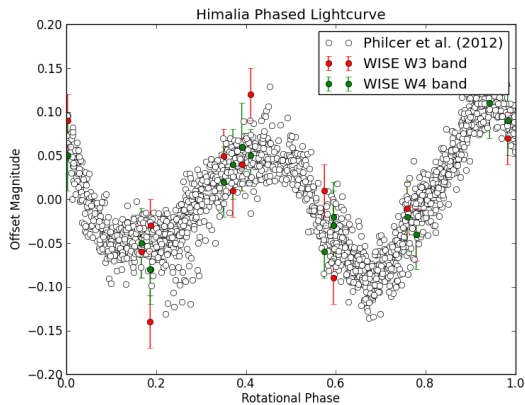}
\caption{The light curve of Himalia from \citet{Pilcher.2012a} and the WISE detections presented in this paper. The WISE detections have been offset in rotational phase by $0.1$ days.}
\label{fig:J6_wise_lc}
\end{center}
\end{figure}

The WISE detections of Himalia show evidence of rotational periodicity. Rotational light curves have been collected of Himalia by \citet{Degewij.1980a} and \citet{Pilcher.2012a}, but with different derived rotational periods. The former collected two nights of observations in November 1976 and reported that the data permitted periods between 9.2 and 9.8 hours with a peak-to-peak amplitude of 0.23 magnitudes in the V band. The latter used multiple nights of observations in 2010 to derive a rotational period of $7.7819\pm0.0005$ hours and an amplitude of $0.20\pm0.01$ magnitudes in the V band. \citet{Pilcher.2012a} noted that \citet{Degewij.1980a} apparently phased their light curve to 2.5 rotational cycles per day, while assuming that three cycles in the interval was the correct way to combine their two nights.  The WISE light curve covers slightly more than one day and has a peak-to-peak amplitude of $~0.2$ magnitude in the W3 band. Figure \ref{fig:J6_wise_lc} shows the light curve derived in \citet{Pilcher.2012a} with the observations from WISE in the W3 (red) and W4 (green) bands overlaid. We find that our observations are offset in rotational phase by $\sim 0.1$ day, which is consistent with the quoted error of $\pm0.0005$ hours considering there are $\sim 190$ rotations between the epochs of the two data sets . The shorter rotational period of \citet{Pilcher.2012a} means that the spectrophotometric observations collected by \citet{Jarvis.2000a} are measurements of the same side of Himalia, rather than sampling of opposite sides as claimed in their paper.

\subsection{J7 Elara}
J7 Elara was discovered in 1905 by Perrine at Lick Observatory. It is in a prograde orbit similar to that of J6 Himalia and is thought to be part of the Himalia dynamical family. While it is the second brightest of the Jovian irregular satellites, little is known about its physical characteristics. \citet{Cruikshank.1977a} used radiometric observations in the $20\mu$m band to derive a diameter of $80\pm20$km and an albedo of $3\pm1\%$ for Elara. \citet{Degewij.1980a} reported UBV colors of Elara, which combined with the low albedo classified the object as C-type. \citet{Degewij.1980a} also noted that their observations in 1975 deviated by as much as 0.5 magnitudes, which they interpreted as possibly due to rotational variation. \citet{Rettig.2001a} also observed Elara using BVR filters, and it was again found to have colors consistent with a C-type surface. They, however, derived an absolute magnitude $H_V = 9.45\pm 0.02$ that was more than 0.6 magnitudes brighter than the $H_V$ reported by \citet{Degewij.1980a}. 

WISE detected Elara in all four bands in the fully cryogenic (four band) portion of the mission, and it was again detected in the W1 and W2 bands during the post-cryogenic phase. The thermal modeling of Elara results in an effective diameter of $79.9\pm1.7$km, with a dark surface albedo of $4.6\pm0.7$\%. Elara has infrared albedos of $6.3\pm0.5\%$ and $5.2\pm1.1\%$ in the W1 and W2 bands respectively. This flat spectrum identify Elara as a C- or P-type surface according to the formalism put forth in \citet{Grav.2012b}. This is consistent with the C-type classification done by other authors \citep{Tholen.1984a,Rettig.2001a}. The low beaming value of $0.79\pm0.03$ indicates that the body has significant surface roughness. 

The WISE detections cover more than 24 hours, and show only $\sim 0.2$ magnitude peak-to-peak in the W3 band rotational lightcurve, significantly less than the variation seen by \citet{Degewij.1980a}. The low peak-to-peak is more in line with the results of \citet{Luu.1991a} which reported a $\sim 0.1$ magnitude variation. \citet{Luu.1991a} speculated that the difference could be explained by a simple change in aspect angle. Jupiter's orbital period is 12 years, allowing a satellite with an obliquity of 90 degrees to change from an "equator on" geometry as seen from Earth to a "pole on" geometry in 3 years. 

\subsection{J8 Pasiphae}
J8 Pasiphae is the third largest of the Jovian irregular satellites and was discovered in 1908 by Philibert~J.~Melotte at the Royal Greenwich Observatory. It orbits the planet in a retrograde orbit with mean inclination of $\sim151^\circ$ and a semi-major axis of $23.6$ million km (or $\sim338$ Jupiter radii), more than twice as distant as J6 Himalia and J7 Elara. Little is known of the physical parameters of this body,  but it has been classified as another C-type asteroid based on photometric measurements \citep{Tholen.1984a,Brown.2000a,Rettig.2001a}.

Detections of J8 Pasiphae was only collected in three of the four WISE bands (W1, W3 and W4) during the fully cryogenic part of the WISE mission. Additional W1 observations were collected during the post-cryogenic phase. The thermal modeling of Pasiphae resulted in an effective diameter of $57.8\pm0.8$km, with a geometric visible albedo of $4.4\pm0.6$\%. With the W1 observations we were able to derive an infrared albedo of $6.7\pm0.7$\% in that band. The relative flat spectrum, assuming a featureless spectra between the visible and infrared, confirms earlier results that the object is C-type. Pasiphae, like J7 Elara, has a low beaming value of $0.76\pm0.02$, indicating it is another body with significant surface roughness. 

Pasiphae show no evidence of significant variation in flux over the 1.2 days spanned by the WISE detections. This confirms the observations by \citet{Luu.1991a}, who also saw minimal light curve amplitude variation in their observations. Again this is in contrast to the results of \citet{Tholen.1984a} and \citet{Degewij.1980a} which discussed the possibility of a rotational variation of as large as one magnitude. This discrepancy could also be due to a change in the aspect angle of the satellite when observed at the different observational epochs.

\subsection{J9 Sinope}
J9 Sinope was discovered in 1914 by Seth~B.~Nicholson at Lick Observatory. It is in a retrograde orbit with mean inclination of $\sim 158^\circ$ and a semi-major axis of $23.9$ million km (or $342$ Jupiter radii). While it is retrograde, it is unlikely to be associated with either the Ananke-Pasiphae complex or the Carme group \citep{Grav.2003b}. It is most likely the largest fragment of a progenitor that was broken up and may also have created Aoede and S/2003 J2, two Jovian irregular satellites that have very similar orbits to that of Sinope. Little is known about the physical properties of this moon, but photometric measurements by \citet{Tholen.1984a} showed that its spectra was different than the other large Jovian irregular satellites they observed. This was confirmed in \citet{Grav.2003b,Grav.2004a}, which showed that Sinope is a D-type object, with spectral slope much steeper than the other "classical" Jovian irregular satellites.  

J9 Sinope was detected in the three of the WISE bands (W1, W3 and W4). The thermal modeling revealed a body with an effective diameter of $35.0\pm0.6$km with a beaming parameter of $0.82\pm0.02$. The geometric visible albedo was found to be $4.2\pm0.6$\%, with a W1 albedo of $10.8\pm1.2$\%. Following the results from \citet{Grav.2012b} this indicates that the Sinope is a D-type object. This is consistent with the results of \citet{Grav.2004a}, which showed that Sinope had colors in the visible and near-infrared that showed a D-type spectral slope. Sinope is another of the largest Jovian irregular satellites that have low beaming value of $0.82\pm0.02$, indicative of low thermal inertia and some surface roughness. 

The WISE detections show a magnitude difference of $\sim 0.1$ in both the W3 and W4 bands for the duration of the observations ($\sim 27$ hours) and no rotational period is evident in our data. The low light curve amplitude is consistent with the results reported in \citet{Luu.1991a}, which gave the rotational period as $13.16$ hours with a peak-to-peak amplitude of $\sim0.2$ magnitudes.

\subsection{J10 Lysithea}
J10 Lysithea was also discovered by Seth Nicholson in 1938 using the Mount Wilson Observatory. It is a prograde orbit similar to J6 Himalia, whose family it belongs too. Almost nothing is known about the physical characteristics of this object, which is thought to be C-type from optical and near-infrared photometric measurements \citep{Grav.2003b,Grav.2004b}. 

Lysithea was detected 15 times during the fully cryogenic phase of the survey, with clear detections in both the W3 and W4 bands in all images. Only 8 of the W1 band images revealed any detectable source. Thermal modeling derived an effective diameter of $42.2\pm0.7$km, with a beaming parameter of $0.93\pm0.02$. The visible albedo fit revealed another dark surface at $3.6\pm0.6$\% reflectivity, which increases to $6.9\pm1.1$\% reflectivity in the W1 band. It is noted that with only about half of the W1 bands having detectable sources, we are most likely sampling only the brightest part of the less than $0.2$ magnitude peak-to-peak rotational light curve of this object in this band. This would lead to the derived W1 albedo possibly being higher than if all parts of the light curve had been sampled. The low W1 albedo, however, identify the object as a C-/P-type object, regardless of this sampling issue.  

\subsection{J11 Carme}
J11 Carme was also discovered by Nicholson in 1938 at Mount Wilson Observatory. Another retrograde orbiting moon, its orbit has a mean inclination of $\sim164^\circ$ and a semi-major axis of 23.4 million km (or $\sim 334$ Jupiter radii). It is the largest of the Carme group of objects, a dynamical family of 19 known irregular satellites with similar orbital elements. 

Carme is the fourth largest of the Jovian irregular satellites and was detected in 16 images in the fully cryogenic phase of the survey and 8 images in the post-cryogenic phase. It was clearly detected in both thermal wavelengths, W3 and W4. In the reflected wavelengths, detections were made in all W1 images, but only in half of the W2 images. Using the  thermal modeling we found a diameter of $46.7\pm0.9$km, with a beaming parameter of $0.95\pm0.03$. The geometric visible albedo is $3.5\pm0.6$\%, which increases significantly to $9.7\pm1.0$\% in the W1 band and $18.5\pm2.2$ in the W2 band. It is cautioned here that W2 albedo might be artificially high as the images missing detections in this band are those corresponding to the trough of the $\sim0.2$ peak-to-peak magnitude rotational light curve in the other bands. The steepness of the spectrum from visible to near-infrared wavelengths indicates a D-type surface, which is consistent with the results of \citet{Grav.2003b,Grav.2004a}.  

\subsection{J12 Ananke}
It would take Nicholson another 13 years after the discovery of J10 Lysithea and J11 Carme, before he discovered his next irregular satellite of Jupiter. J12 Ananke was first seen in 1951, again using Mount Wilson Observatory. Its orbit has a mean inclination of $\sim 148^\circ$ and a semi-major axis of $\sim21.3$ million km (or $\sim304$ Jupiter radii). Ananke is therefore part of the Pasiphae-Ananke complex, a large group of irregular satellite with similar orbital elements that may be the result of one or two fragmenting events \citet{Grav.2003b}. Optical and near-infrared photometry showed that this object has a P-type surface \citep{Grav.2003b,Grav.2004a}.

Ananke was only detected in the two thermal bands, W3 and W4, with 15 flux measurements in each band. This allowed for a thermal model solution of $29.1\pm0.6$km with a beaming value of $1.01\pm0.03$. The geometric visible albedo shows another dark surface with $3.8\pm0.6$\% reflectivity. The object appears to have a rotational light curve with a peak-to-peak variation of $\sim0.4$ magnitudes. 

\subsection{J13 Leda}
Leda, the smallest of the 8 "classical" Jovian irregular satellites, was discovered in 1974 by Charles T. Kowal using the Mount Palomar Observatory. It belongs to the prograde Himalia group and is thought to be another C-type surface based on optical and infra-red photometric measurements \citep{Grav.2003b,Grav.2004a}.  Leda was detected by WISE in the W3 and W4 bands only . Thermal modeling using the 12 detections in each of the band yields a diameter of $21.5\pm1.7$km with a beaming parameter of $1.15\pm0.13$. The geometric visible albedo is again very dark, at $3.4\pm0.6$\% reflecticity. 

\subsection{J17 Callirrhoe}
This retrograde Jovian irregular satellite was discovered in 1999, by the Spacewatch project. It was orginally designated as an asteroid (1999 UX18), but was identified as an irregular satellite of Jupiter the following year and given the designation S/1999 J1. Callirrhoe became the 17th confirmed moon of Jupiter. With an inclination of $\sim 147^\circ$ Callirrhoe may be part of the Pasiphae cluster, but it might also be part of its own dynamical complex outside the Pasiphae-Ananke complex. Little is known about its physical propertied, but it has been as having a D-type surface \citep{Grav.2003b, Grav.2004a}. 

There were 10 detections of this object in the W3 band and 8 in the W4 band, which allowed for a thermal model fit using beaming as a free parameter. The effective diameter was found to be $9.6\pm1.3$km with a beaming parameter of $0.85\pm0.17$. With a H value of $13.92\pm0.02$ from \citet{Grav.2003b} the visible albedo was found to be $5.2\pm1.6$\%. 

\subsection{J23 Kalyke}
J23 Kalyke was discovered in 2000 by Scott Sheppard using telescopes at Mauna Kea Observatory in Hawaii. It has an orbit similar to that of Carme, and if therefore thought to be in the Carme dynamical family. \citet{Grav.2003b} used optical photometry to classify Kalyke as a D-type object. While Kalyke passed through 14 individual images during the fully cryogenic WISE phase, it was not detected in any of them. However, stacking these exposures revealed a $\sim4\sigma$ source at the predicted position in the W3 band. No source was found in the corresponding W4 band stack. 

The measured magnitude in the W3 band is $11.32\pm0.24$ using the method laid out in \citet{Wright.2010a} using aperture photometry. Using an assumed beaming value of $0.9\pm0.2$ yielded an effective diameter of $6.9\pm1.3$km. The H value of $15.28\pm0.04$ from \citet{Grav.2003b} was used to derive a geometric visible albedo of $2.9\pm1.4$.

\subsection{J27 Praxidike}
J27 Praxidike is another Jovian irregular that was discovered by Sheppard in 2000. It is in a retrograde orbit with an orbit similar to that of the objects in the Ananke family. Optical photometry shows that Praxidike has a C-type surface, which is inconsistent with the Ananke family having homogeneous surface throughout its membership \citep{Grav.2003b,Grav.2004a}.

There were 13 images covering this smaller irregular satellite, but detections were not found in any of the individual exposures. The stacked image in the W3 band, using all 13 exposures, revealed a $\sim4\sigma$ source at the predicted position. No source was seen in the W4 band in the corresponding stack. The measured stacked magnitude in the W3 band, at $11.43\pm0.28$, was used together with an assumed beaming value of $0.9\pm0.2$ to derive an effective diameter of $7.0\pm1.4$km. Using the H value of $15.24\pm0.03$ from \citet{Grav.2003b} we derive a geometric visible albedo of $2.9\pm0.6\%$, another dark surfaced Jovian irregular satellite. 

\subsection{S9 Phoebe}
S9 Phoebe, discovered by W. H. Pickering in 1899 at Boyden Observatory, was the first retrograde moon to be discovered in the solar system and is by far the most comprehensive studied object among the irregular satellites. It was observed by ground based telescopes in the 1970's, and it was found to have brightness fluctuations of 0.2 mag with a period of either 11.3 or 21.5 hours, with rather gray UBV colors\citep{Andersson.1974a}. The satellite was also found to have a very high phase coefficient of 0.10-0.15 mag/deg, which led \citet{Cruikshank.1979a} and \citet{Degewij.1980a,Degewij.1980b} to conclude that Phoebe appeared to be similar to a C-type asteroid. \citet{Thomas.1983a} used 50 disk resolved observations of Phoebe from the Voyager 2 spacecraft to derive a mean diameter of $220\pm20$km and a rotational period of $9.4\pm0.2$ hours. This yielded a geometric albedo of $6.9\pm2\%$. 
	
\citet{Kruse.1986a} performed photometric observations of Phoebe and improved the determination of the rotational period to $9.282\pm0.015$ hours. They also improved the mean opposition V magnitude, which revised the geometric albedo to $8.4\pm0.3\%$ when using the \emph{Voyager 2} derived diameter. \citet{Kruse.1986a} also derived a strong opposition effect of $0.180\pm0.035$ mag/deg, which they showed was consistent with that of the C-type asteroids.  \citet{Simonelli.1999a} later used the same images from \emph{Voyager 2} to construct an albedo map of Phoebe, showing most of its surface having normal reflectance between $7$ to $10\%$. 

\citet{Bauer.2004a} provided new observations of the rotational period of Phoebe, improving it to $9.2735\pm0.0006$ hours, allowing for the correlation of previously observed spectral features, colors, and albedo feature with observations of the moon by \emph{Cassini} during its 2004 June 11 encounter. This flyby resulted in a unprecedented view of an irregular satellite for slightly more than three rotations, with resolution to better than 2 km per pixel across the surface \citep{Porco.2005a}. It was found that the moon has a mean radius of $106.6\pm1$ and a mean density of $1630\pm45$ km per cubic meter, which for a porosity of $<\sim40\%$ requires a mixture of ice and rock of some type. 

WISE detected Phoebe in all four bands, although the detections in the $4.6\mu$m are low signal-to-noise. The thermal fit yields a diameter of $202.2\pm4.5$ with an albedo of $10.0\pm0.5\%$. While this diameter is slightly lower than that derived by Cassini, the two results are within $2\sigma$ of each other. The derived beaming value of $1.15\pm0.03$ is also on the high end for the irregular satellites observed reported here, with only J13 Leda having similarly high beaming value. 


\subsection{S26 Albiorix}
S26 Albiorix is the largest of the Gallic cluster of prograde irregular satellites of Saturn, centered on $34^\circ$ mean inclination. It was discovered in by a team lead by Brett Gladman in 2000 and was given the designation S/2000 S11 \citep{Gladman.2001a}. Optical and near-infrared photometry revealed it as a P-type \citep{Grav.2003b,Grav.2004a}.

Albiorix was another satellite that were not seen in individual images, but was revealed using shift-and-stack of 11 available W4 frames. The resulting detection has a signal-to-noise ratio of $\sim3$. Thermal modeling yields a diameter of $28.6\pm5.4$km with a visible albedo of $6.3\pm2.7$, when assuming a default value of $0.9\pm0.2$ for the beaming parameter. 

\subsection{S29 Siarnaq}
S29 Siarnaq is  prograde satellite and the largest in the Inuit group centered on $45^\circ$ mean inclination. It was also discovered in 2000 by a team lead by Gladman \citep{Gladman.2000a}. This object was classified as a P-type object through optical and infrared broadband photometry \citep{Grav.2003b,Grav.2004a,Buratti.2005a}. While there is little physically known about the satellite, it appears  to be in a secular resonance with Saturn \citep{Nesvorny.2003a,Cuk.2004a}. 

WISE did not detect Siarnaq in individual images, but when shift-and-stack of the 13 available frames was performed in W4 it revealed a detection with signal-to-noise ratio of $\sim7$. Thermal modeling yields a diameter of $39.3\pm5.9$km and visible albedo of $5.0\pm1.7$, assuming a beaming value of $0.9\pm0.2$. 

\section{W1 and W2 Albedo}
Seven of the irregular satellites detected by WISE were bright enough to be detected in the W1 and/or W2 band. For these distant, cold objects, these two bands, centered on $3.4$ and $4.6$ microns, are dominated by reflected light. This allows for the determination of "near-infrared" albedo in these two bands. \citet{Grav.2012b} found a correlation between the W1 albedo and taxonomic type for the Jovian Trojan population, with the objects having $< 9\%$ albedo in W1 being correlated with C- and P-type asteroids and those with higher W1 albedos correlating with D-type asteroids. This correlation is continued in this data set. The five objects (Himalia, Elara, Pasiphae, Lysithea and Phoebe) with $p_{3.4} < 9\%$ are all C-type objects \citep{Grav.2003b}. The other two objects (Sinope and Carme) with $p_{3.4} > 9$ are both D-type objects \citep{Grav.2003b}. 

\section{Discussion}
The irregular satellites, particularly the retrograde Jovian irregulars, show some of the lowest visible geometric albedos on any population of small solar system bodies observed by WISE, with a inverse variance weighted mean albedo of $4.3\pm1.4$\%.  With the exception of S9 Phoebe, every irregular satellite observed by WISE is on the dark end of the distribution of geometric visible albedo found in the dark component of the main asteroid belt, the Hilda population, the Jovian Trojan population, the blue component of the Centaur population, and the cometary population. \citep{Masiero.2011a,Grav.2011b,Grav.2012a,Grav.2012b,Bauer.2013a}. The low albedo of the irregular satellites may be indicative of an origin for these bodies among the other dark solar system populations, or may trace unique physical or evolutionary processes that these bodies were subjected to between formation and the present day. Further photometric and taxonomic studies of these objects will enable us to place them in context of the rest of the solar system and help to constrain potential evolutionary pathways. 

Both S9 Phoebe and the other Saturnian irregular satellites have been invoked to explain as a source region of the dark material found on the leading side of the Saturnian regular satellite S6 Iapetus \citep{Soter.1974a,Buratti.2005a}. Numerical simulations predict that the irregular satellites underwent a significant collisional evolution that may have generated large quantities of dark dust \citep{Nesvorny.2003a, Turrini.2009a}, which was collaborated with the recent discovery and confirmation of a dust ring originating from Phoebe \citep{Verbiscer.2009a}. However, data collected using the \emph{Cassini} Visual and Infrared Mapping Spectrometer (VIMS) during the flybys of Phoebe and Iapetus showed that few spectral associations exist between the two objects \citep{Tosi.2010a}. While Phoebe might not be a dominant source of the dark material, the significantly redder surfaces of S26 Albiorix and S29 Siarnaq \citep{Grav.2003b, Grav.2007a} remain as possible candidates. The low visible albedos found for these two objects strengthens that possibility, but their prograde orbits means that dust from these objects have much lower likelihood of colliding with Iapetus than dust from retrograde irregular satellites. \citet{Tamayo.2011a} studied the dynamics of dust in the Saturnian system and found that dust from the retrograde Saturnian irregular satellites can have comparable likelihoods of ending up on Iapetus as those coming from Phoebe. They argued that this mix of Phoebe's flat spectral slope and the redder spectral slope of the other retrograde irregular satellites could contribute to the differences in the spectra between Phoebe and Iapetus. Unfortunately, WISE did not observe any of the other retrograde Saturnian irregular satellites, but the confirmation of the dark surfaces of Albiorix and Siarnaq (as well as the uniformly dark surfaces of the Jovian irregular satellites) in this study strengthens the belief that the other irregular satellites of Saturn could have similar dark surfaces.

\section{Conclusion}
We present thermal model fits for 11 Jovian and 3 Saturnian irregular satellites based on measurements from the WISE dataset. Our fits confirm spacecraft-measured diameters for the objects with {\it in situ} observations (Himalia and Phoebe) and provide diameters and albedo for 12 previously unmeasured objects. The best-fit thermal model beaming parameters are comparable to what is observed for other small bodies in the outer Solar System, while the visible, W1, and W2 albedos trace the taxonomic classifications previously established in the literature. Reflectance properties for the irregular satellites measured are similar to the Jovian Trojan and Hilda Populations, implying common origins. The irregular satellites, particularly the retrograde Jovian irregulars, show some of the lowest visible geometric albedos on any population of small solar system bodies, with a inverse variance weighted mean albedo of $4.3\pm1.4$\%.

\section{Acknowledgments}
This publication makes use of data products from the {\it Wide-field Infrared Survey Explorer}, which is a joint project of the University of California, Los Angeles, and the Jet Propulsion Laboratory/California Institute of Technology, funded by the National Aeronautics and Space Administration. This publication also makes use of data products from NEOWISE, which is a project of the Jet Propulsion Laboratory/California Institute of Technology, funded by the Planetary Science Division of the National Aeronautics and Space Administration. We gratefully acknowledge the extraordinary services specific to NEOWISE contributed by the International Astronomical Union's Minor Planet Center, operated by the Harvard-Smithsonian Center for Astrophysics, and the Central Bureau for Astronomical Telegrams, operated by Harvard University. We also thank the worldwide community of dedicated amateur and professional astronomers devoted to minor planet follow-up observations. This research has made use the NASA/IPAC Infrared Science Archive, which is operated by the Jet Propulsion Laboratory/California Institute of  Technology, under contract with the National Aeronautics and Space Administration. 

We also thank Fredrick Pilcher, of Organ Mesa Observatory in Las Cruces, New Mexico, for providing us with the light curve data of JVI Himalia from his paper \citep{Pilcher.2012a}.

\end{document}